# Bose-Einstein Condensation of Excitons in Bilayer Electron Systems

J.P. Eisenstein[1] & A.H. MacDonald[1,2]

[1]*Dept. of Physics, California Institute of Technology, Pasadena, CA 91125 USA;* [2]*Dept. of Physics, University of Texas, Austin, TX 78712 USA.*

*'These authors contributed equally to this work'*

**An ordered state of electrons in solids in which excitons condense was proposed many years ago as a theoretical possibility but has, until recently, never been observed. We review recent studies of semiconductor bilayer systems that provide clear evidence for this phenomenon and explain why exciton condensation in the quantum Hall regime, where these experiments were performed, is as likely to occur in electron-electron bilayers as in electron-hole bilayers. In current quantum Hall exciton condensates, disorder induces mobile vortices that flow in response to a supercurrent and limit the extremely large bilayer counterflow conductivity.**

In many-particle quantum physics bosons are special. Unlike the ubiquitous electrons and their fermionic cousins, any number of bosons can crowd into the same microscopic state. Indeed, in 1924 Einstein predicted that at low temperatures essentially *all* of the bosons in a macroscopic system would spontaneously "condense" into the same low energy quantum state.

In quantum mechanics, an individual particle is represented by a wave that has both amplitude and phase. The most remarkable consequence of Bose-Einstein condensation (BEC) of a vast number of particles is that macroscopic properties become dependent on a single wavefunction, promoting quantum physics to classical length and time scales. A BEC is a highly ordered state in which the wavefunction phase is coherent over distances much longer than the separation between individual particles. Long-range quantum phase coherence has many dramatic physical consequences of which the most spectacular is superfluidity, *i.e.* the ability of matter to flow around obstacles with extremely weak, often immeasurable, dissipation.

All known BECs are formed by particles which are actually composite bosons made up of an even number of fermions. For example, helium-4 atoms, which contain two protons, two neutrons, and two electrons, Bose condense and form a superfluid at just a few degrees above absolute zero. Today there is intense interest in BECs in rarefied atomic vapors, where over 100 fermions are often present inside each composite bosonic atom. In superconductors electrons form pairs at low temperatures. These "Cooper pairs" are again composite bosons. In this case, the formation of pairs and their collective Bose condensation usually occur at the same low temperature; the transition temperature is controlled by the underlying fermionic physics not purely by bosonic quantum mechanics. Correspondingly, the average separation between electrons in a given Cooper pair usually greatly exceeds the mean distance between pairs, in sharp contrast to the essentially point-like bosons present in liquid helium or the atomic vapor BECs.

An exciton in a semiconductor consists of an electron bound to a hole. In most cases the electron is in the conduction band of the semiconductor while the hole, which is nothing more than an unfilled electronic state masquerading as a positively charged object, is in the valence band. Excitons are usually created by shining light on the semiconductor, which creates electrons and holes in equal numbers. Optically generated excitons like these are ephemeral objects, decaying quickly via the emission of light.

Electrons and holes are both fermions, but excitons are bosons. After the 1957 pairing theory of superconductivity was solidly established, physicists began to speculate[1,2] about BECs involving excitons in semiconductors. The short lifetime of optically generated excitons has proven to be a major obstacle in the search for such a system[3,4]. Indirect excitons in bilayer quantum well systems have two key advantages over bulk systems and have thus played an increasingly important role[5,6]. Quantum wells are realized in layered semiconductor structures and allow for the confinement of electrons and holes to two-dimensional planes. Indirect excitons are bound states of conduction band electrons in one well and valence band holes in an adjacent well. The spatial separation between the electrons and holes reduces the rate at which the optically generated particles decay back into photons. In addition, owing to this spatial separation these excitons act like small dipoles and therefore repel one another. This repulsion is important because it prevents the electrons and holes from agglomerating into an uninteresting electron-hole plasma. Experimental studies of optically generated indirect excitons have uncovered a number of fascinating many-body effects[7,8,9,10]. The enhanced exciton mobility, increased radiative decay rate, and photoluminescence noise that has been reported in these systems at low temperatures and high fields, all suggest increased coherence under conditions that are expected to be favorable for exciton condensation. Still, the requirements of high density and low temperatures are difficult to achieve in this non-equilibrium system, and there is not yet convincing evidence for an exciton BEC.

In this Progress Article we describe recent compelling evidence that exciton condensation has been discovered where it has been least expected, in systems with two parallel layers of conduction band electrons. Importantly, the excitons in this system are present in equilibrium, waiting patiently for experimenters to reveal their properties.

Surprisingly, the discovery we discuss here has come in bilayer *electron-electron* systems, like the ones illustrated schematically in Figure 1. In such systems the Fermi level lies in the conduction band in both layers. The techniques required to grow high-quality single and bilayer electron systems are now well established and have been key to a great many important physics discoveries, notably including the famous fractional quantum Hall effect[11]. At first sight it seems impossible to achieve BEC of excitons in electron-electron bilayers, because all the fermions have repulsive interactions. Quantum well electrons are, however, able to perform remarkable tricks when placed in a strong perpendicular magnetic field.

The Lorentz force bends classical electron trajectories into circles – cyclotron orbits. In a two-dimensional quantum system the kinetic energy of these orbits is quantized into discrete units. The set of orbits with a particular energy, a *Landau level*, is however highly degenerate since the tiny cylcotron orbits can be positioned all across the 2D plane. It turns out that the number of degenerate states in the lowest energy Landau level is equal to the number of quanta of magnetic flux that pass through the electron layer. At high field it is easy to enter a regime in which the total number of electrons is less than the number of states in the lowest Landau level. In Figure 1, for example, we illustrate the circumstance in which the number of electrons in each layer is 1/3 of the number of available states; this is referred to as filling factor $\nu = 1/3$.

BEC at strong magnetic fields in electron-electron bilayers is most easily understood by making a particle-hole transformation in one of the two layers; we have chosen the bottom layer for this in Figure 1. The particle-hole transformation[12,13] is one in which we simply keep track of the empty states in the Landau level rather than the full ones. It is mathematically exact, changes the filling factor of the transformed layer from $\nu$ to $1-\nu$, and changes the sign of the carrier charge from negative to positive. The interaction of holes with electrons in the un-transformed layer is thus attractive.

Particle-hole transformations also change the sign of the kinetic energy, from negative to positive for valence band holes at zero magnetic field for example. Because the kinetic energy is the same for all electrons and holes in the lowest Landau level, BEC at strong fields is just as likely to occur in electron-electron bilayers as in electron-hole bilayers. This point is a subtle one; indeed spontaneous coherence[14] and superfluidity[15] were predicted in electron-electron bilayers without recognizing the equivalence of earlier predictions[16] for electron-hole bilayers. We emphasize, however, that particle-hole transformation of a conduction band Landau level is completely equivalent to the much more familiar transformation used to map unoccupied valence band electron states into holes.

In Fig. 1 the number of holes in the bottom layer exceeds the number of electrons in the top layer. In order to create an exciton BEC, these populations should be nearly equal; one should therefore begin with *half-filled* Landau levels in each layer. Early experiments proved that in this 1/2+1/2 situation a bilayer electron system can exhibit a quantized Hall effect[17]. In other words, its Hall resistance, measured with electrical currents flowing in parallel through the two layers, is precisely equal to $h/e^2$, Planck's constant divided by the square of the electron charge. This remarkable effect results from a complex interplay of Landau quantization, Coulomb interaction effects, and imperfections in the 2D plane. While its observation in the bilayer system at this filling factor demonstrates the importance of interlayer Coulomb interactions, it does not on its own suggest the existence of an exciton BEC. The hunt for BEC requires different experimental tools.

In a bilayer 2DES the two quantum wells are separated by a thin barrier layer. By adjusting the thickness and composition of this barrier, it is possible for electrons in one layer to interact strongly, via the Coulomb interaction, with the electrons in the other layer and yet have very little probability of quantum mechanically tunneling through the barrier. This strong correlation - weak tunneling limit is key to the existence of excitonic Bose condensation in the bilayer 2DES. Ironically, it was measurements of the weak residual inter-layer tunneling rate in a bilayer 2DES that provided the first compelling evidence for exciton condensation[18].

Figure 2 shows the rate at which electrons tunnel between the two layers of a bilayer 2DES as a function of the voltage difference between the layers. These data were taken at very low temperatures and with the magnetic field adjusted to produce the half-filling per layer condition most suitable for exciton condensation. The two traces in the figure differ only in the effective interlayer separation. For the blue trace, the layers are relatively far apart. For voltages near zero, the tunneling rate is very small. This suppression of the tunneling is not sensitive to small changes in the magnetic field and grows more severe as the temperature is reduced. Often referred to as a Coulomb gap, the effect is a consequence of strong correlations between electrons *in the individual layers*. Tunneling at low voltages (or, equivalently, low energies) is suppressed because an electron attempting to enter a 2D electron layer at high magnetic field blunders disruptively into the delicate dance being performed by the electrons in that layer as they skillfully avoid each other. This clumsy entry can only produce a highly excited, non-equilibrium state and can occur only at high voltage. While quite interesting in its own right, the effect obviously does not suggest exciton condensation.

The red trace is Fig. 2 is dramatically different. A slight reduction in the effective layer separation has produced a giant peak in the tunneling rate. In sharp contrast to the suppression effect in the blue trace, this peak grows rapidly as the temperature is reduced below about 0.5K and is very sensitive to the magnetic field. Small changes in the field away from the value needed to produce half-filled lowest Landau levels rapidly destroy the peak and restore the suppression effect.

The stark difference between the two traces in Fig. 2 suggests that a quantum phase transition occurs as a function of effective layer separation. The strong peak in the tunneling rate seen at small separations is inconsistent with the arguments used to understand the suppression effect at larger separations. Instead of being unaware of the correlations in the layer about to be entered, the tunneling electron is apparently already participating in the dance. Indeed, the peak suggests that all electrons are strongly correlated with their neighbors in *both* layers. Moreover, the temperature and magnetic field dependence of the peak point to the existence of a new and intrinsically bilayer collective state in which electrons in one layer are always positioned opposite holes in the other layer.

Strong electron-hole correlations are a necessary but not sufficient condition for exciton condensation. In order to make the argument for excitonic BEC in a bilayer 2D electron system more compelling, an experiment demonstrating the transport of electron-hole pairs is needed. But how does one move and detect neutral objects? The key is to note that the uniform flow of excitons is equivalent to ordinary electrical currents flowing in opposite directions in the two layers. Technical tricks have allowed us to make independent electrical connections to the individual layers in bilayer electron samples[19]. (Indeed, separate contacts were also essential in obtaining the tunneling data in Fig. 1.) With these contacts it is easy to arrange for equal and opposite currents to flow in the two layers and directly test whether or not excitons are available to perform the particle transport .

Figure 3 shows a cartoon version of what is expected from such a counterflow measurement. The two traces represent the Hall voltages expected in the two layers, neglecting all quantum effects save exciton condensation. Because of the Lorentz force on a moving charge, the Hall voltage is, in most cases, simply proportional to the magnetic field. In a bilayer system with counterflowing currents, the Hall voltages in the two layers will be of opposite sign.

If the distance between the two layers is too large, or if the magnetic field is far from the half filling per layer condition, exciton condensation will not occur. If, however, the layers are closely spaced and the magnetic field is just right, interlayer electron-hole pairs will form and carry the counterflow current. *The Hall voltage in each layer should then drop to zero*, as suggested in the figure. This prediction can be understood in a variety of ways, most simply by observing that excitons are neutral and thus feel no Lorentz force. Without the Lorentz force the Hall voltage must vanish. This remarkable prediction has recently been confirmed by our own group at Caltech[20] and the effect has been reproduced, in a slightly different bilayer system, by researchers at Princeton[21]. Interestingly, tiny voltages drops in the direction of current flow are observed in each layer. These voltages, which appear to have an activated temperature dependence, imply that the electron-hole transport current flows with weak but measurable dissipation, much like the supercurrent in the mixed state of superconductors. As we now explain, this similarity is likely more than coincidental.

In quantum mechanics the velocity of a particle is proportional to the rate at which the wavefunction phase changes with position. In an excitonic BEC the same simple description holds; the velocity of a gas of electron-hole pairs is proportional to the spatial gradient of the condensate phase. Because the wavefunction is collective, elementary quantum processes which act on one particle at time cannot effectively reduce the velocity of a BEC. In two-dimensional superfluids condensate current decay is limited instead by the nucleation and transport of vortices, points around which the condensate phase changes by $2\pi$. The elementary process which leads to phase gradient decay is one in which a vortex moves across the system, as illustrated in Figure 4. The energy required to create an isolated vortex in the ground state of a ideal two-dimensional BEC is (logarithmically) infinite. At temperatures below the Kosterlitz-Thouless transition temperature, at which the *free energy* of an isolated vortex vanishes, no unpaired vortices should be present. However, if free vortices are present in a *real* system, their motion will lead to condensate current decay at finite temperatures.

Vortices are especially important in bilayer exciton condensates because they tend to be nucleated by disorder. The special role of vortices in these superfluids can be understood by thinking about the Aharonov-Bohm phases associated with the external magnetic field. As explained above, the number of states N in a Landau level is proportional to the number of quanta of magnetic flux that penetrate the two-dimensional layers. Correspondingly, the total phase change of an electron orbital that encircles the system is $2\pi N$. A vortex in the condensate phase therefore either increases or decreases the number of states that can be accommodated in one of the two-layers by a single unit. Vortices consequently come in four flavors[22] and carry charge equal in magnitude to one half of the electron charge. In bilayer exciton condensates vortices couple to, and are therefore nucleated by, external disorder potentials. The experimental results discussed above demonstrate that vortex motion in the present excitonic BECs requires only a finite activation energy. In all current samples, free vortices are present down to the lowest temperatures studied.

The property that vortices are present, likely down to T=0, suggests analogies between the properties of excitonic BECs at high magnetic field and the properties of type-II superconductors. In the mixed state of superconductors, vortices are nucleated by an external magnetic field and are present in the ground state of the system. Robust, and technologically significant, superconductivity occurs only when the vortex positions are pinned by disorder. This analogy suggests that it might be possible to enhance the superflow properties of excitonic BECs by learning how to pin the vortices, something that has not yet been seriously attempted.

The road from Keldysh's suggestion that there might be an excitonic analog of superconductivity to these recent discoveries has been punctuated by unanticipated twists and turns. Excitonic BEC requires strong inter-band interactions, but very small amplitude for tunneling between occupied bands as a consequence of either external potentials or interactions. The possibility of studying the bilayer quantum wells systems that satisfy this requirement so beautifully had to wait for the development of the techniques we now have for growing extremely high quality layered semiconductors. The interaction energy gained by excitonic condensation must compete with band energies and energy gains associated with other competing types of order. The understandings that Landau quantization in a strong field could eliminate the band energy cost (*even in electron-electron bilayers*), and that excitonic condensation could prevail over other orders under some circumstances, both grew out of the theory of the fractional quantum Hall effect -- something that had not even been anticipated at the beginning of this journey. Finally, the question of how collective transport of neutral objects could be detected in an excitonic BEC has always been a thorny issue. The separate contacting techniques that enable counterflow transport measurements, developed with quite different goals in mind, provide a direct solution to this problem. Keldysh's speculations have proven correct, although not in exactly the way he had in mind. What's more, the presence of vortices in these superfluids makes the analogy to superconductivity more complete than he could have imagined.

This research was supported in part by the National Science Foundation (J.P.E. and A.H.M.) and the Department of Energy (J.P.E.). In addition, A.H.M. acknowledges support from the Moore Scholar program and the hospitality of the California Institute of Technology. We thank Anton Burkov, Yogesh Joglekar, Melinda Kellogg, Loren Pfeiffer, Enrico Rossi, Ian Spielman, and Kenneth West for their essential help in this research.

The authors declare that they have no competing financial interests.

Correspondence and requests for materials should be addressed to J.P.E. (e-mail: jpe@caltech.edu).

**Figure 1**: An electron-electron bilayer system in a strong magnetic field is equivalent to an electron-hole bilayer. **Upper drawing:** Cartoon depiction of two parallel layers of electrons. **Middle drawing:** In a magnetic field the kinetic energy of two-dimensional electrons is quantized into discrete Landau energy levels. Each such Landau level contains a huge number of degenerate orbitals, here depicted schematically as a checkerboard of sites. If the field is strong enough, all electrons reside in the lowest Landau level, and only occupy a fraction (here 1/3) of the available sites. **Lower drawing:** A particle-hole transformation applied to the lower electron layer places the emphasis on the unoccupied sites, *i.e.* the holes, colored green, in that layer. This transformation, which is formally exact and completely equivalent to the more familiar transformation used to describe empty valence band states in a semiconductor as holes, changes the sign of the Coulomb interactions between layers from repulsive to attractive. Exciton BEC occurs when holes in the lower layer bind to electrons in the upper layer. This is most likely when the number of electrons and holes are equal, i.e. when each layer is 1/2 filled. This is not the case in the present figure.

**Figure 2**: Tunneling rate versus interlayer voltage in a bilayer electron system. These traces are actual data, and are taken at magnetic fields where exciton condensation is most expected (*i.e.* one-half filling of the lowest Landau level per layer). In the blue trace the layers are relatively far apart, while in the red trace they are closer together. The dramatic difference between them is a direct indication that a phase transition in the bilayer system occurs when the layer separation is reduced below a critical value. The huge enhancement of the tunneling rate at zero energy in the red trace points to interlayer electron-hole correlations: *i.e*. it suggests that every electron is positioned opposite a hole into which it can easily tunnel.

**Figure 3**: Hall voltage measurements reveal exciton condensation. The two traces schematically indicate the Hall voltages in the two electron layers when the electrical currents flowing in them are oppositely directed. All quantum effects, save exciton condensation, have been ignored. When the currents are carried by *independent* charged particles in the two layers, non-zero Hall voltages must be present to counteract the Lorentz forces. Since the currents are oppositely directed, these voltages have opposite signs in the two layers. However, if exciton condensation occurs at some magnetic field, the oppositely directed currents in the two layers are carried by a uniform flow of excitons in one direction. Being charge neutral, these excitons experience no Lorentz force and the Hall voltage is expected to vanish. This remarkable effect has very recently been definitively observed.

**Figure 4**: Motion of unpaired vortices leads to dissipation in excitonic superfluids. The elementary excitations in quantum Hall excitonic superfluids are vortices which carry electrical charge and topological charge. In an ideal system at low temperatures vortices occur in bound pairs and do not contribute to the decay of excitonic supercurrents. Disorder, however, is present in all real samples and is capable of producing unpaired vortices. The weak dissipation observed in recent counter-flow experiments is associated with the activated transport of such unpaired vortices.

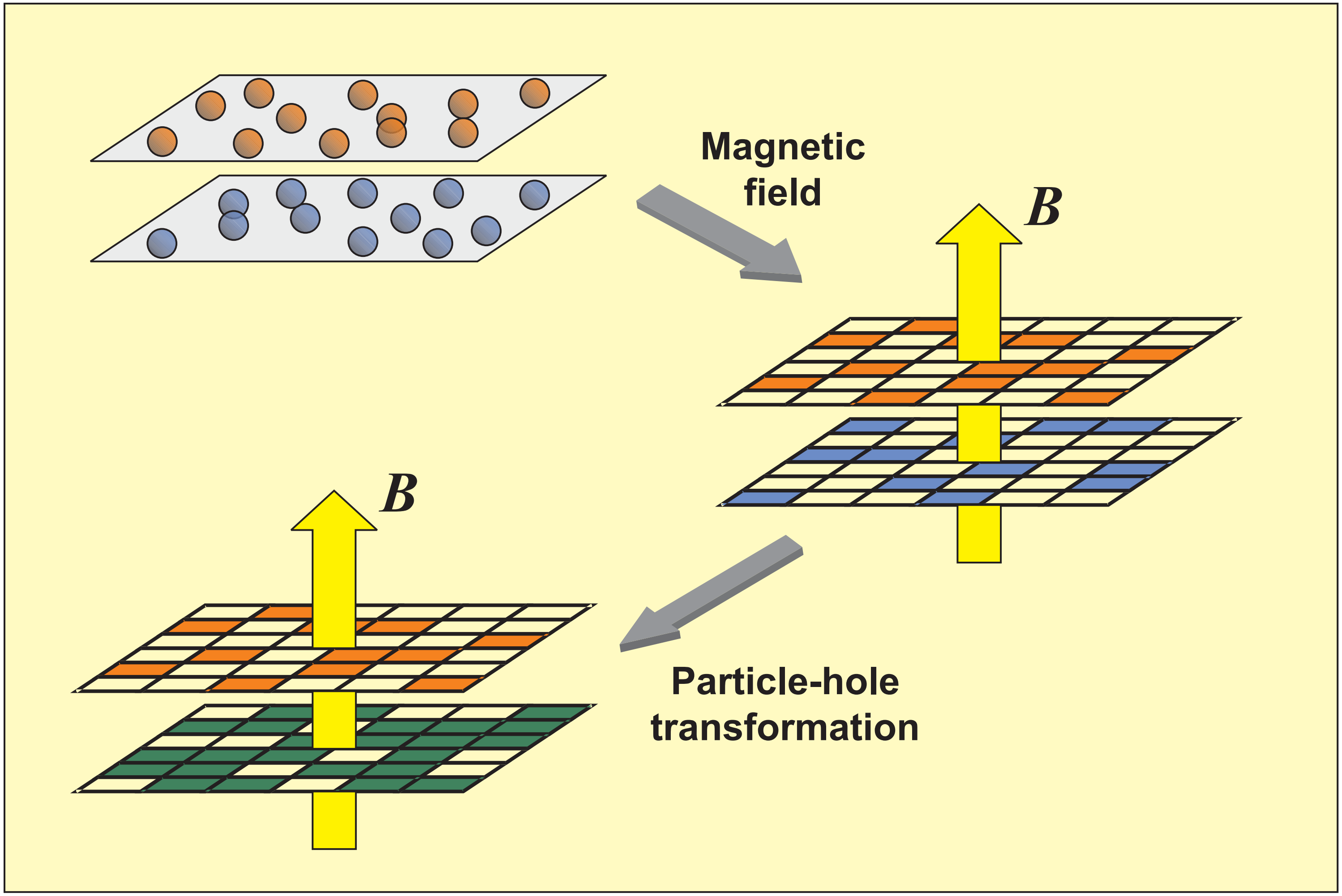

Figure 1

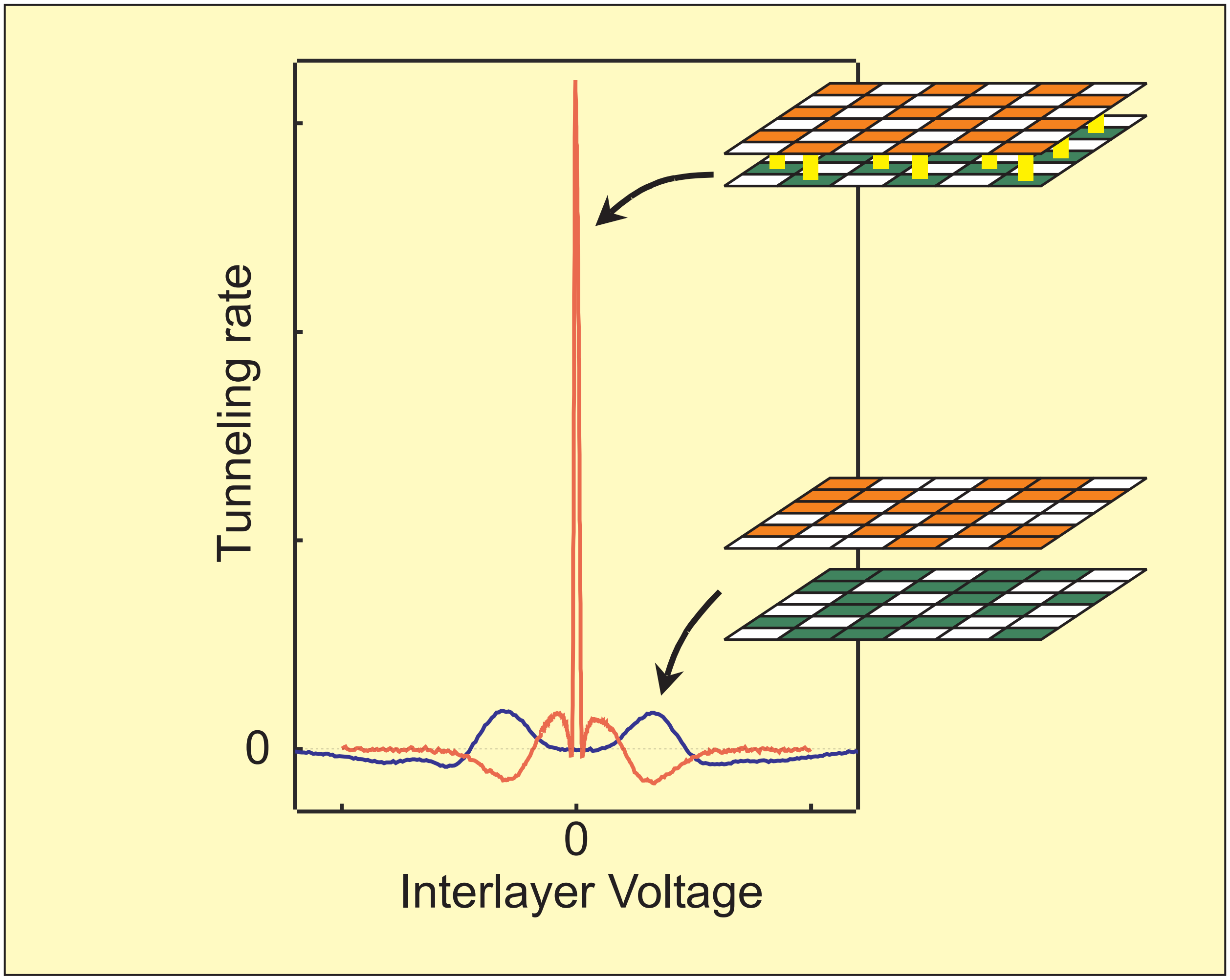

Figure 2

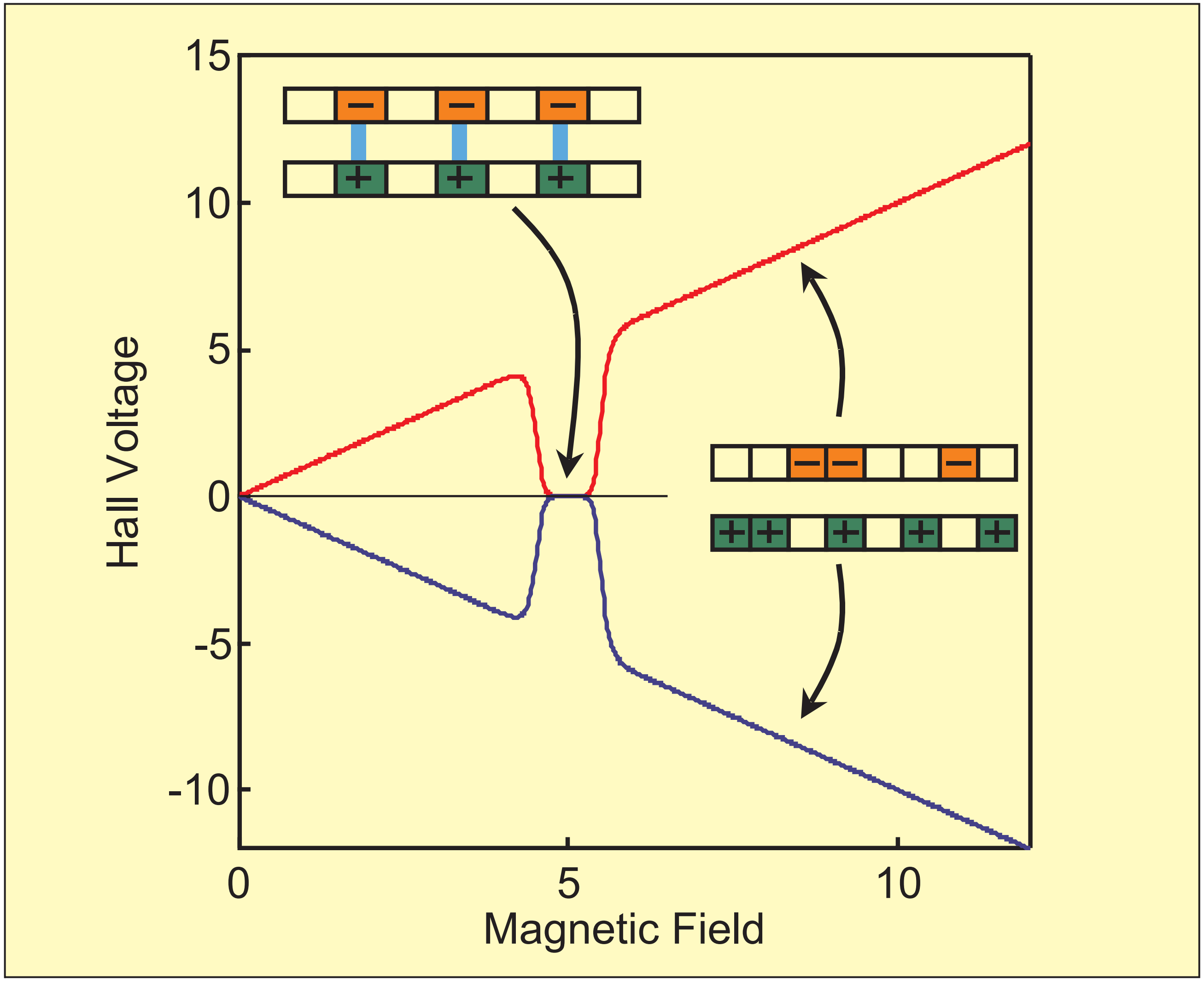

Figure 3

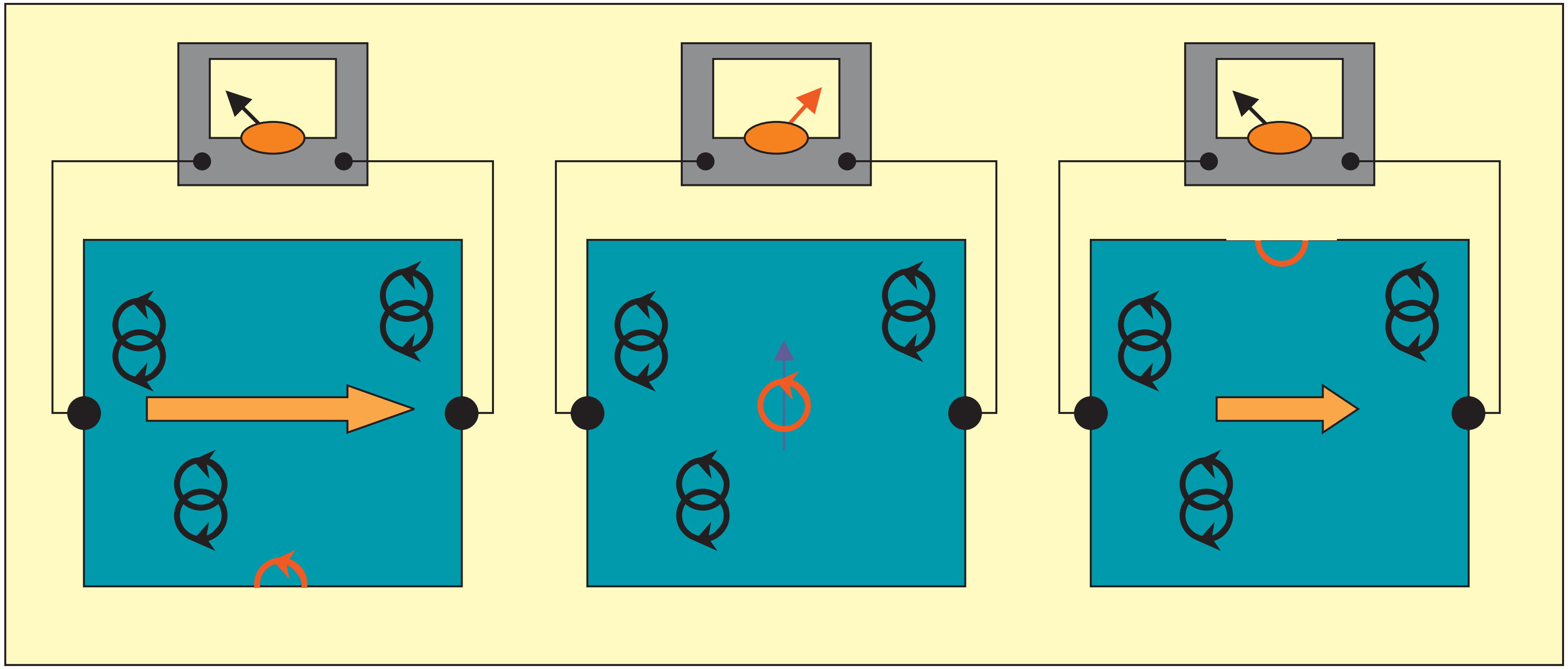

Figure 4